\documentclass[journal]{vgtc}                

\ifpdf
  \pdfoutput=1\relax                   
  \usepackage{graphicx}                
  \DeclareGraphicsExtensions{.pdf,.png,.jpg,.jpeg} 
\else
  \usepackage{graphicx}                
  \DeclareGraphicsExtensions{.eps}     
\fi%

\graphicspath{{figures/}{pictures/}{images/}{./}} 

\usepackage{microtype}                 
\PassOptionsToPackage{warn}{textcomp}  
\usepackage{textcomp}                  
\usepackage{mathptmx}                  
\usepackage{times}                     
\usepackage{cite}                      
\usepackage{tabu}                      
\usepackage{booktabs}                  

\usepackage{color}
\usepackage{enumitem}
\usepackage{wrapfig}
\usepackage{amssymb}
\usepackage{amsmath}

\definecolor{my_purple}{rgb}{0.73, 0.39, 0.81}
\definecolor{my_blue}{rgb}{0.41, 0.62, 1.0}
\definecolor{my_green}{rgb}{0.0, 0.53, 0.21}

\newcommand{\name}[1]{GAN Lab}

\preprinttext{Published in IEEE Transactions on Visualization and Computer Graphics}

\onlineid{0}

\vgtccategory{Research}
\vgtcpapertype{application/design study}

\title{GAN Lab: Understanding Complex Deep Generative Models using Interactive Visual Experimentation}

\author{Minsuk Kahng, Nikhil Thorat, Duen Horng (Polo) Chau, Fernanda B. Vi\'{e}gas, and Martin Wattenberg}
\authorfooter{
\item
 Minsuk Kahng and Duen Horng (Polo) Chau are with Georgia Institute of Technology. E-mail: \{kahng\,$|$\,polo\}@gatech.edu.
\item
 Nikhil Thorat, Fernanda B. Vi\'{e}gas, and Martin Wattenberg are with Google Brain. E-mail: \{nsthorat\,$|$\,viegas\,$|$\,wattenberg\}@google.com.
}

\shortauthortitle{Kahng \MakeLowercase{\textit{et al.}}: GAN Lab: Understanding Complex Deep Generative Models using Interactive Visual Experimentation}

\renewcommand{\manuscriptnotetxt}{This paper will be published in the IEEE Transactions on Visualization and Computer Graphics (TVCG), Vol. 25, No. 1, January 2019.}

\abstract{
Recent success in deep learning has generated immense interest among practitioners and students, inspiring many to learn about this new technology.
While visual and interactive approaches have been successfully developed to help people more easily learn deep learning, most existing tools focus on simpler models.
In this work, we present \name{}, the first interactive visualization tool  designed for non-experts to learn and experiment with \textit{Generative Adversarial Networks (GANs)}, a popular class of complex deep learning models.
With \name{}, users can interactively train generative models
and visualize the dynamic training process's intermediate results. 
\name{} tightly integrates an \textit{model overview graph} that summarizes GAN's structure, and a \textit{layered distributions} view that helps users interpret the interplay between submodels.
\name{} introduces new interactive experimentation features for learning complex deep learning models, such as \textit{step-by-step} training at multiple levels of abstraction for understanding intricate training dynamics.
Implemented using \textit{TensorFlow.js}, \name{} is accessible to anyone via modern web browsers, without the need for installation or specialized hardware, overcoming a major practical challenge in deploying interactive tools for deep learning.

} 

\keywords{Deep learning, information visualization, visual analytics, generative adversarial networks, machine learning, interactive experimentation, explorable explanations}

\CCScatlist{ 
 \CCScat{H.5.2}{Information Interfaces and Presentation}{User Interfaces}
}

\teaser{
\centering
\includegraphics[width=\linewidth]{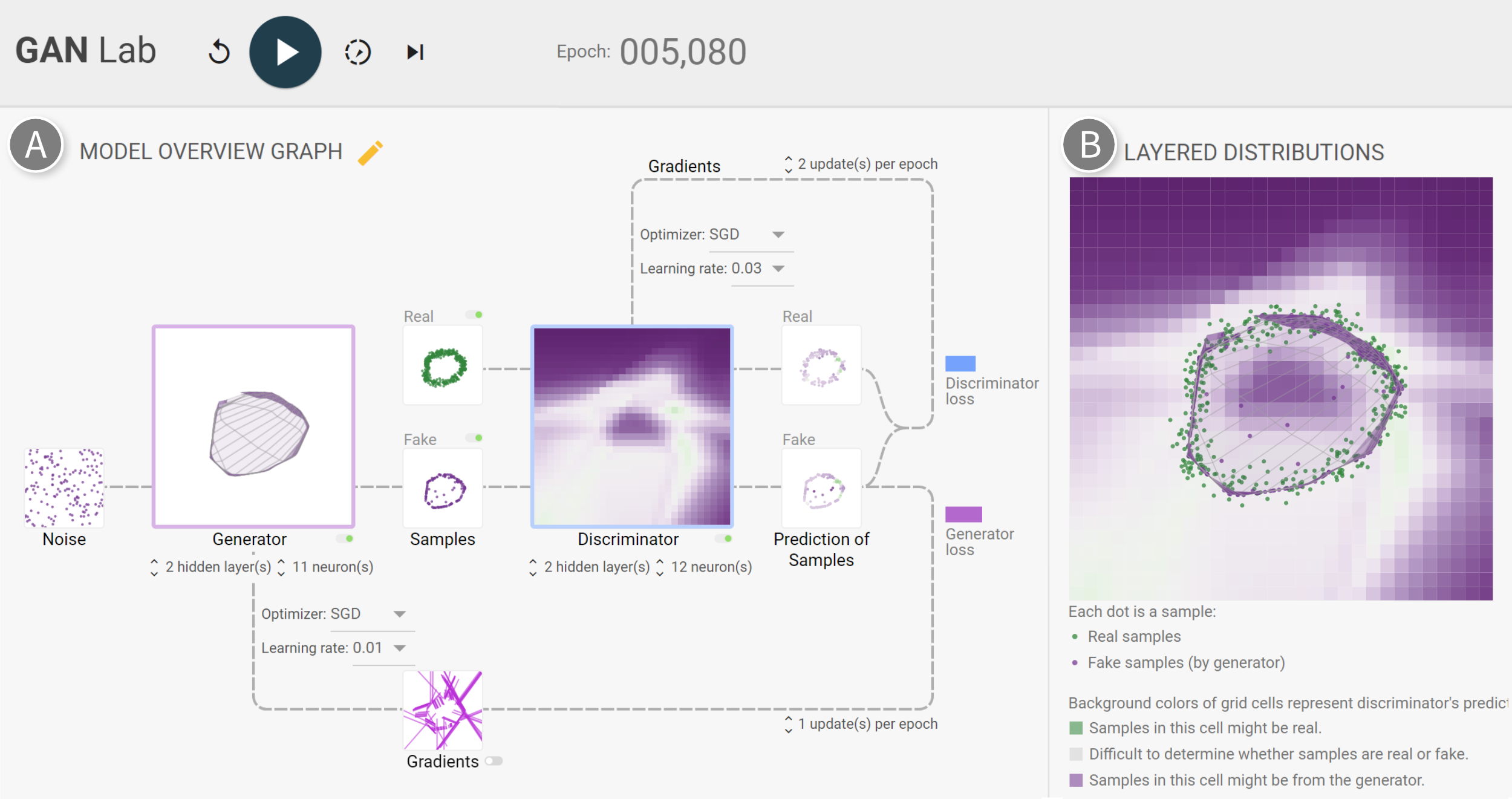}
\vspace{-7mm}
\caption{With \name{}, users can interactively train Generative Adversarial Networks (GANs), and visually examine the model training process.
In this example, a user has successfully used \name{} to train a GAN that generates 2D data points whose challenging distribution resembles a ring.
\textbf{A.} The \textit{model overview graph} summarizes a GAN model's structure as a graph, with nodes representing the \textcolor{my_purple}{generator} and \textcolor{my_blue}{discriminator} submodels, 
and the data that flow through the graph (e.g., fake samples produced by the generator).
\textbf{B.} The \textit{layered distributions} view 
helps users interpret the interplay between submodels through user-selected layers, such as the discriminator's classification heatmap, \textcolor{my_green}{real samples}, 
and \textcolor{my_purple}{fake samples} produced by the generator.
}
\label{figure:teaser}

}

\vgtcinsertpkg

\begin{document}

\firstsection{Introduction}

\maketitle

\label{sec:intro}

Recent success in deep learning has generated a huge amount of interest from practitioners and students,
inspiring many to learn about this technology.
Visual and interactive approaches have successfully been used to describe concepts and underlying mechanisms in deep learning~\cite{olah2014neural,karpathymnist,smilkov2016directmanipulation,yosinski2015understanding}.
For example,
Karpathy's popular interactive demo~\cite{karpathymnist} enables users to run convolutional neural nets and visualize neuron activations, inspiring researchers to develop more interactive tools for deep learning.
Another notable example is Google's \textit{TensorFlow Playground}~\cite{smilkov2016directmanipulation}, an interactive tool that visually represents a neural network model and allows users to interactively experiment with the model 
through direct manipulation;
Google now uses it to educate their employees about deep learning~\cite{2018mlcc}. 

\noindent \textbf{The rise of GANs and their compelling uses.}
Most existing interactive tools, however, have been designed for simpler models. 
Meanwhile, modern deep learning models are becoming more complex.
For example, \textit{Generated Adversarial Networks (GANs)}~\cite{goodfellow2014generative}, 
a class of deep learning models
known for their remarkable ability to
generate synthetic images that look like natural images,
are difficult to train and for people to understand, even for experts.
Since the first GAN publication by Goodfellow et al.~\cite{goodfellow2014generative} in 2014, 
GANs have become one of the most popular machine learning research topics~\cite{2017ganzoo,lecun2016what}.
GANs have achieved state-of-the-art performance in a variety of previously difficult tasks, such as 
synthesizing super-resolution images based on  low-resolution copies, and performing image-to-image translation (e.g., converting sketches to realistic images)~\cite{goodfellow2016nips}.

\begin{figure}[b]
 \centering
 \includegraphics[width=\columnwidth]{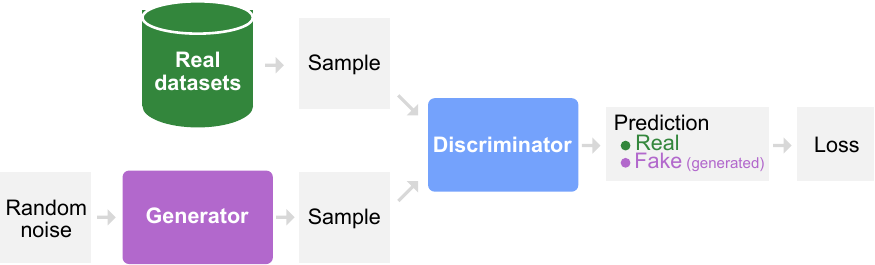}
 \vspace{-5mm}
 \caption{A graphical schematic representation of a GAN's architecture commonly used.}
 \label{figure:architecture-diagram}
\end{figure}

\vspace{2pt}
\noindent 
\textbf{Key challenges in designing learning tools for GANs.}
At the high level, a GAN internally combines two neural networks, called \textit{generator} and \textit{discriminator}, 
to play a game where the generator creates  ``fake'' data and the discriminator guesses 
whether that data is real or fake
(both types of data are mixed together).
A perfect GAN is one that generates fake data that is virtually indistinguishable from real data.
A user who wishes to learn about GANs needs to develop a mental model of not only what the two submodels do, 
but also how they affect each other in its training process.
The crux in learning about GANs, therefore, originates from the iterative, dynamic, intricate interplay between these two  submodels.
Such complex interaction is challenging for novices to recognize, and sometimes even for experts to fully understand~\cite{salimans2016improved}.
Typical architecture diagrams for GANs (e.g., \autoref{figure:architecture-diagram}, commonly shown in learning materials) do not effectively help people develop the crucial mental models needed for understanding GANs.

\begin{figure}[tb]
 \centering
 \includegraphics[width=\linewidth]{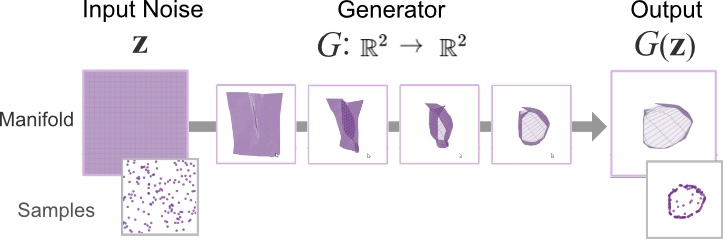}
 \vspace{-5mm}
 \caption{
 In \name{}, the \textit{generator}'s non-trivial data transformation is visualized as a manifold, which turns \textit{input noise} (leftmost) into fake samples (rightmost). 
 \name{} animates the input-to-output transformation to help users more easily understand this complex behavior.
 }
 \label{figure:generator-transformation}
\end{figure}

\vspace{3pt}
\noindent \textbf{Contributions.} 
In this work, we contribute:

\begin{itemize}[topsep=0mm, itemsep=0mm, parsep=1mm, leftmargin=3mm]

\item \textbf{\name{}, the first interactive tool designed for non-experts} to learn and experiment with GAN models, a popular class of complex deep learning models, that overcomes multiple unique challenges for developing interactive tools for GANs (\autoref{sec:challenges}).

\item \textbf{Novel interactive visualization design} of \name{} (\autoref{figure:teaser}), 
which tightly integrates a \textit{model overview graph} that summarizes GAN's structure (\autoref{figure:teaser}A) as a graph, selectively visualizing components that are crucial to the training process; 
and a \textit{layered distributions} view (\autoref{figure:teaser}B) that helps users interpret the interplay between submodels through user-selected layers (\autoref{sec:tool}).
\name{}'s visualization techniques work in tandem to help crystalize complex concepts in GANs.
For example, \name{} visualizes the generator's data transformation, which turns input noise into fake samples, 
as a manifold (\autoref{figure:teaser}, big box with purple border).
When the user hovers over it, 
\name{} animates the input-to-output transformation (\autoref{figure:generator-transformation}) to visualize
how the input 2D space is folded and twisted by the generator to create the desired ring-like data distribution,
helping users more easily understand the complex behavior of the generator.

\item \textbf{New interactive experimentation features} for learning complex deep learning models, such as  \textit{step-by-step}  training at multiple levels of abstraction for understanding intricate training dynamics (\autoref{sec:experimentation}). The user can also interact with the training process by directly manipulating GAN's hyperparameters.

\item \textbf{A browser-based, open-sourced implementation} that helps \textit{broaden public's education access to modern deep learning technologies} (\autoref{sec:implementation}).
Training deep learning models conventionally requires significant computing resources. 
For example, deep learning frameworks, like TensorFlow~\cite{abadi2016tensorflow}, typically run on dedicated servers. 
They are not designed to support low-latency computation needed for real-time interactive tools,
or large number of concurrent user sessions through the web.
We overcome such practical challenges in deploying interactive visualization for deep learning by using \textit{TensorFlow.js Core},\footnote{TensorFlow.js {\small (\url{https://js.tensorflow.org})} was formerly deeplearn.js.} 
an in-browser GPU-accelerated deep learning library 
recently developed by Google;
the second author is a lead developer of TensorFlow.js Core.
Anyone can access \name{}  using their web browsers without the need for installation or specialized backend.
\name{} runs locally on the user's web browser, 
allowing us to easily scale up deployment for our tool to the public, significantly broadening people's access to tools for learning about GANs.
The source code is available in \url{https://github.com/poloclub/ganlab/}.

\item Usage scenarios that demonstrate how \name{} can help beginners learn key concepts and training workflow in GANs, and assist practitioners to interactively  attain optimal hyperparameters for reaching challenging equilibrium between submodels (\autoref{sec:scenario}).

\end{itemize}

\vspace{5pt}
\noindent \textbf{VIS's central role in AI.}
We believe in-browser interactive tools developed by our VIS community, like \name{}, will play critical roles in promoting people's understanding of deep learning, and raising their awareness of this exciting new technology.
To the best of our knowledge, our work is the first tool designed for non-experts to learn and experiment with complex GAN models,
different from recent work in visualization for deep learning~\cite{liu2017towards,strobelt2018lstmvis,liu2018analyzing,kahng2018activis,pezzotti2018deepeyes,wongsuphasawat2018visualizing} 
which primarily targets machine learning experts. 
Our work joins a growing body of research that aims to use interactive visualization to explain complex inner workings of modern machine learning techniques.
Distill, a new interactive form of journal, is dedicated to achieving this exact goal~\cite{olah2017research}.
We hope our work will help inspire even more research and development of visualization tools that help people better understanding artificial intelligence technologies.

\section{Background: Generative Adversarial Networks}

This section presents a brief introduction of Generated Adversarial Networks, which will help ground our discussion in this paper.

Generative Adversarial Networks (GANs)~\cite{goodfellow2014generative} are a new class of unsupervised generative deep learning models that
 model data distributions.
It can be used for generating multi-dimensional data distributions (e.g., an image is a multi-dimensional data point, where each pixel is a dimension).
The model takes \textit{real samples} and random vectors (i.e., \textit{random noise}) as inputs and transforms the random vectors into \textit{fake samples} that mimic the real samples.
Ideally, the distribution of the fake samples will be indistinguishable from the real samples.
The architecture of GANs is composed of two neural networks, called \textit{generator} and \textit{discriminator}, and
is often represented as an abstracted data-flow graph as in \autoref{figure:architecture-diagram}.
The generator, $G$, takes a random noise vector, $\mathbf{z}$, as input and transforms it into a fake sample, $G(\mathbf{z})$ (i.e., a multi-dimensional vector); 
the discriminator, $D$, which is a binary classifier, takes either a real or fake sample, and determines whether it is real or fake ($D(\mathbf{x})$ represents the probability that $\mathbf{x}$ is real rather than fake).

A GAN model is iteratively trained through a game between the discriminator and generator.
In GAN, two cost functions exist:
the one for the discriminator measures the probability of assigning the correct labels to both real and fake samples (i.e., the sum of $D(\mathbf{x})$ and $1 - D(G(\mathbf{z}))$);
the other for the generator measures that for fake samples only (i.e., $1 - D(G(\mathbf{z}))$).
The goal of the discriminator is to maximize its cost, but the goal of the generator is to minimize its cost,
which introduces conflicts (i.e., zero-sum). Therefore, it has to play a mini-max game to find the optimum.
Goodfellow et al.~\cite{goodfellow2014generative} used an interesting analogy to explain how it works, 
where we can view the generator as a \textit{counterfeiter} who makes fake dollar bills, 
and the discriminator as the \textit{police}. 
If the police can spot the fake bills, that means the counterfeiter is not ``good enough,''
so the counterfeiter carefully revises the bills to make them more realistic.
As the discriminator (police) differentiates between real and fake samples, 
the generator (counterfeiter) can glean useful information from the discriminator to revise its generation process so that it will generate more realistic samples in the next iteration.
And to continue to receive such helpful information, the generator keeps providing its updated samples to the discriminator.
This iterative interplay between the two players leads to generating realistic samples.

\section{Related Work}

\subsection{Visualization for Understanding Deep Learning}

Researchers and practitioners have written articles and deployed explorable web-based demos to help people learn about concepts in deep learning.
One of the popular examples is Chris Olah's series of essays,\footnote{Colah's blog, {\small \url{http://colah.github.io}}}
explaining mathematical concepts behind deep learning using visualizations.
One article explains how neural networks transform and manipulate manifolds~\cite{olah2014neural}. 
Another popular example is Andrej Karpathy's collection of web-based demos developed using ConvNetJS,\footnote{ConvNetJS, {\small \url{https://cs.stanford.edu/people/karpathy/convnetjs/}}} a lightweight JavaScript library for deep learning.
His MNIST demo~\cite{karpathymnist}
dynamically visualizes intermediate results, such as neuron activation.

Olah's articles and Karpathy's demos have inspired many researchers to develop interactive visualizations for novices to easily understand deep learning techniques~\cite{smilkov2016directmanipulation,harley2015isvc}.
A notable example is \textit{TensorFlow Playground}~\cite{smilkov2016directmanipulation}, an interactive visualization tool for non-experts to train simple neural net models.
Google has integrated it into its internal machine learning course for educating its employees;
the course is now available to the public~\cite{2018mlcc}.
Distill, a new online interactive journal, has recently been created and it is dedicated to interactive explanation of machine learning~\cite{olah2017research}.
The journal features a growing number of articles with interactive visualization~\cite{wattenberg2016how,goh2017why,carter2017using}.
However, 
most existing visualizations focus on simpler models.
Modern deep learning models are much more complex,
and we will present and discuss unique design challenges that stem from such complexity (\autoref{sec:challenges}).

\subsection{Algorithm Visualization \& Explorable Explanations}

Even before the surge of interest in deep learning techniques, researchers had studied how to design interactive visualization to help learners better understand the dynamic behavior of algorithms~\cite{shaffer2010algorithm,hundhausen2007you,saraiya2004effective,hundhausen2002meta}.
These tools often graphically represent data structures and allow students to execute programs in a step-by-step fashion~\cite{shaffer2010algorithm,guo2013online}.
While many of these tools target algorithms covered in undergraduate computer science curricula,
some specialized tools exist for artificial intelligence~\cite{amershi2005designing}.
As deep learning models are a category of specialized algorithms, when we design \name{}, we draw inspiration from the principles and guidelines proposed in the aforementioned related domains~\cite{schweitzer2007interactive}.

As web has become a central medium for sharing ideas and documents,
many interactive experimentation tools implemented in JavaScript have been viewed as ``explorable explanations,''
an umbrella term coined by Bret Victor in 2011~\cite{victor2011explorable}.
He advocated the use of interactive explanations with examples to help people better understand complex concepts by actively engaging in the learning process.
Many interactive tools instantiate this idea,
including the ones showcased on the popular website with the same name (Explorable Explanations\footnote{Explorable Explanations, \small\url{http://explorabl.es/}}).
These tools aim to help people actively learn through playing and interactive experimentation.
\name{} aligns with this research theme.

\subsection{Visual Analytics for Deep Learning Models \& Results}

Over the past few years,
many visual analytics tools for deep learning have been developed~\cite{liu2017towards,strobelt2018lstmvis,ming2017understanding,kahng2018activis,liu2018analyzing,wang2018ganviz,bilal2018convolutional}, as surveyed in \cite{hohman2018visual,liu2017towardsb}.
Most were designed for experts
to analyze models and their results.
For instance,
TensorFlow Graph Visualizer~\cite{wongsuphasawat2018visualizing} visualizes model structures, to help researchers and engineers build  mental models about them.
Many other tools 
focus to visually summarize model results for interpreting how models respond to their datasets.
For example, CNNVis~\cite{liu2017towards} was designed for inspecting CNN model results;  LSTMVis~\cite{strobelt2018lstmvis} and RNNVis\cite{ming2017understanding} were for RNN models.
A few other tools allow users to diagnose models during training.
For example, DeepEyes~\cite{pezzotti2018deepeyes} does so  through t-SNE visualizations.
Two visual analytics tools have been developed for GANs~\cite{liu2018analyzing,wang2018ganviz}.
DGMTracker~\cite{liu2018analyzing} allows experts to diagnose and monitor the training process of generative models through visualization of time-series data on data-flow graphs.
GANViz~\cite{wang2018ganviz} helps experts evaluate and interpret trained results through multiple views, including one showing the distributions of real and fake image samples, for a selected epoch, using t-SNE.
Different from all existing tools designed to help experts analyze models and results that we summarized above,
we focus on non-experts and learners, helping them build intuition of the internal mechanisms of models, through interactive experimentation.

\section{Design Challenges for Complex Deep Learning Models}
\label{sec:challenges}

\begin{figure*}[t]
 \centering
 \includegraphics[width=\linewidth]{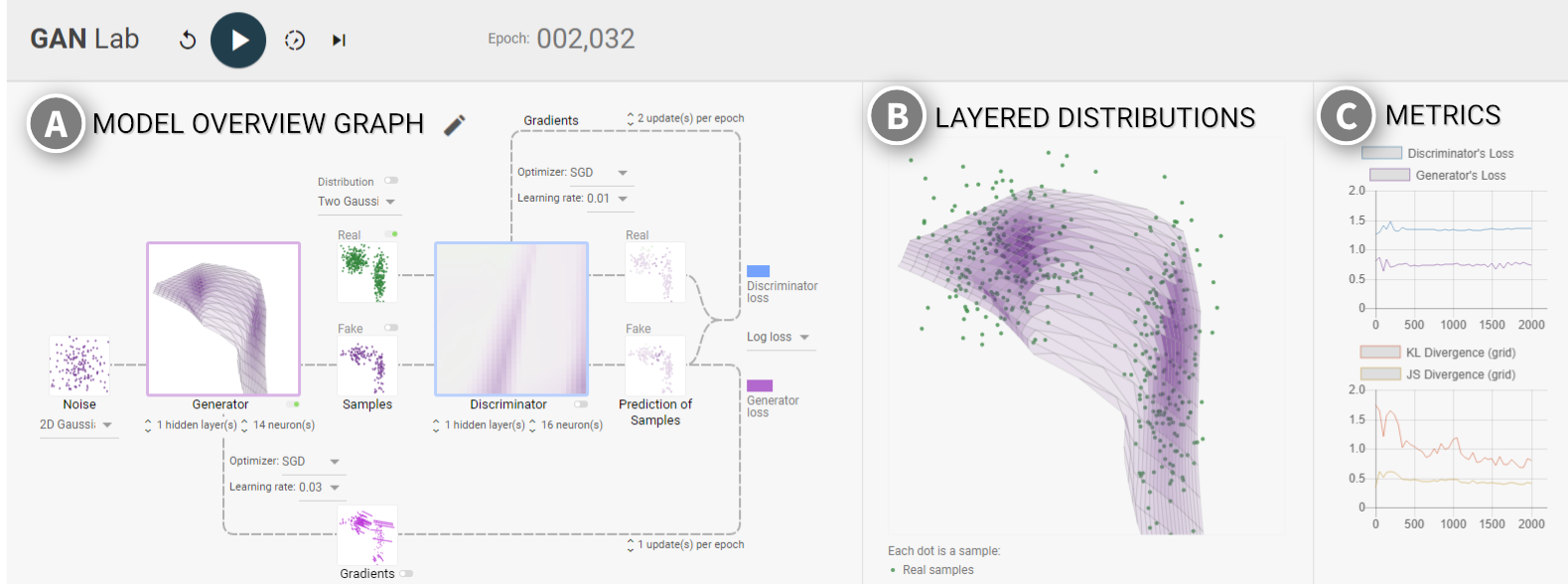}
 \vspace{-5mm}
 \caption{The \name{} interface integrates multiple views: 
 \textbf{A.} The \textit{model overview graph} summarizes a GAN model's structure as a graph, with nodes representing the submodels, and the data that flow through the graph;
 \textbf{B.} The \textit{layered distributions} view overlays magnified versions of the graph's component visualizations, 
 to help users more easily compare and understand their relationships;
 \textbf{C.} The \textit{metrics} view presents line charts that track metric values over the training process.
 Users start the model training by clicking the \textit{play} button on menu bar.
 The three views are dynamically updated, as training  progresses.
 In this example, real samples are drawn from two Gaussian distributions, and the generator, consisting of a single hidden layer with 14 neurons, has created samples whose distribution is quite similar to that of the real samples.
 }
 \label{figure:screenshot}
\end{figure*}

Our goal is to build an interactive, visual experimentation tool for users to better understand GANs, a complex deep learning model.
To design \name{},
we identified four key design challenges unique to GANs.

\begin{enumerate}[label=C\arabic*.,itemsep=1mm, topsep=2mm]

\item {[\textsc{Model}]} \textbf{Complex model structures with submodels.}
The structures of modern deep learning models (including GANs) are complex;
they often incorporate multiple base neural networks or deep learning models as submodels.
For example, a GAN combines two neural nets: generator and discriminator;
an image captioning model often consists of both CNNs and RNNs 
for translation between images and text~\cite{vinyals2015show}.
Effective visualization of such models calls for new strategies different from those designed for conventional models.
For example, it is crucial to find the appropriate levels of visual abstraction for the models, as
visualizing all low-level details will overwhelm users.
Special visual design may be needed to help users interpret 
the intricate interplay between submodels (e.g., discriminator and generator).

\item {[\textsc{Data}]} \textbf{
High-dimensional datasets.}
As deep learning models often work with large, high-dimensional datasets,
visualizing their distributions would quickly create many traditional challenges well-studied in information visualization research~\cite{liu2015visualizing}.
While we may use techniques like \textit{dimensionality reduction} to partially address such issues, 
this could introduce additional complexities to the systems, potentially distracting users from their
main goal of understanding how deep learning models work.

\item {[\textsc{Training Process}]} \textbf{Many training iterations until convergence.}
Deep learning models are trained through many iterations (i.e., at least thousands),
introducing nontrivial challenges for developing interactive  tools.
First of all, as it takes time to converge, 
the tools need to keep providing users with information
during training (e.g., progress), and
users may also want to provide feedback to models 
(e.g., by changing hyperparameters).
In addition,
one popular feature used in many experimentation tools is a step-by-step execution of systems~\cite{guo2013online,saraiya2004effective},
however, the definition of \textit{steps} becomes different in training of complex models,
because the training process consists of many iterations and each iteration also consists of the training of multiple submodels.

\item {[\textsc{Deployment}]} 
\textbf{Conventional deep learning frameworks ill-fitted for multi-user, web-based deployment.}
Training deep learning models conventionally requires significant computing resources.
Most deep learning frameworks written in Python or C++, like TensorFlow~\cite{abadi2016tensorflow}, typically run on dedicated servers that utilize powerful hardware with GPU, to speed up the training process.
However, even with a powerful backend, they cannot easily support a large number of concurrent user sessions through the web, 
because each session requires significant computation resources.
When combined, even a small number of concurrent sessions can bog down a powerful server.
Off-loading computation to the end user is a possible solution, but conventional deep learning frameworks 
are not designed to support low-latency computation needed for real-time interactive tools.

\end{enumerate}

\section{Design Goals}

Based on the identified design challenges in the previous section,
we distill the following main design goals 
for \name{}, a novel interactive visualization tool for learning and experimenting with GANs.

\begin{enumerate}[label=G\arabic*.,itemsep=1mm, topsep=2mm]

\item \textbf{Visual abstraction of models and data flow.}
To give an overview of the structure of complex models,
we aim to create a visual representation of a model by selectively choosing and grouping low-level operations (and intermediate data) into high-level components (C1).
It helps users visually track how input data are transformed throughout the models.
For users to clearly examine the internal model training process and data flow, 
we would use low-dimensional datasets (C2).
(\autoref{sec:architecture})

\item \textbf{Visual analysis of interplay between discriminator and generator.}
As GANs internally use two different neural nets,
it is important for users to understand how they work together, to get a holistic picture of the overall training process (C1).
In response, we would like to enable users to examine and compare the visualizations of the model components to understand they affect each other to accomplish the generation tasks.
(\autoref{sec:layered})

\item \textbf{Dynamic experimentations through direct manipulation of hyperparameters.}
We aim to let users dynamically play and experiment with models.
To help users quickly understand the roles of many hyperparameters and control them (C3),
we would like to design interactive interfaces which users can easily locate and manipulate the options. The users' actions are directly applied to the model training process.
(\autoref{sec:manipulate-options})

\item \textbf{Supporting step-by-step execution for learning the training process in detail.}
Since the training process of deep learning models consists of many iterations and each iteration also consists of several steps, 
the step-by-step execution of models can greatly help novices to understand the training process (C3).
To address this needs, we aim to design multiple ways to execute models in a step-by-step fashion by decomposing the training process into steps at multiple levels of abstraction.
(\autoref{sec:step-all})

\item \textbf{Deployment using cross-platform lightweight web technologies.}
To develop a tool that is accessible from multiple users without a need to use specialized powerful backend (C4),
we would like to use web browsers both for training models and visualizing results.
(\autoref{sec:implementation})

\end{enumerate}

\section{Visualization Interface of \name{}}
\label{sec:tool}

This section describes \name{}'s interface and visualization design.
\autoref{figure:screenshot} shows \name{}'s interface, consisting of multiple views.
Using the control panel on top, users can run models and control the speed of training, which we describe in detail in the next section (\autoref{sec:experimentation}).
This section primarily describes the other three views that visualize models and trained results:
(A) \textbf{model overview graph} view on the left (\autoref{sec:architecture});
(B) \textbf{layered distributions} view in the middle (\autoref{sec:layered});
(C) \textbf{metrics} view on the right (\autoref{sec:metrics}).
In the figure, 2D real samples are drawn from two Gaussian distributions. 
The user's goal is to train the model so that it will generate a similar distribution, by transforming 2D Gaussian noise using a neural net with a single hidden layer.

\textbf{Color scheme.}
In our visualization, we color real data \textcolor{my_green}{green} and fake data \textcolor{my_purple}{purple}. 
We do not use a more traditional green-red color scheme, as we do not want to associate fake data with a negative value.
For visualizing the discriminator,
we use \textcolor{my_blue}{blue}, a color unrelated to the color scheme chosen for coloring data. 
For visualizing the generator,
we again use the color \textcolor{my_purple}{purple} because the generated points are the fake points the model sees.

\subsection{Model Overview Graph: Visualizing Model Structure and Data Flow}
\label{sec:architecture}

The \textit{model overview graph} view (\autoref{figure:screenshot} at A) visually represents a GAN model as a graph, by selectively grouping low-level operations into high-level components and presenting data flow among them.

\subsubsection*{Abstraction of Model Architecture as Overview Graph}

The model overview graph visually summarizes the architecture of a GAN model.
Instead of presenting all low-level operations and intermediate data (i.e., output tensors),
it selectively represents high-level components and important intermediate data as nodes.
Specificallly, nodes of the graph include 
two main submodels (i.e., generator and discriminator) and several intermediate data (e.g., fake samples).
Each submodel, which is a neural network, is represented as a large box, 
and six data nodes are visualized as small boxes.
This decision is based on our observation of how people draw the architecture of GANs~\cite{creswell2018generative} (like \autoref{figure:architecture-diagram}).
Users are often familiar with the structure of the basic neural networks and more interested in the overall picture and interplay between the two submodels.
we place input data nodes on the left side of the submodels and output nodes on the right (for forward data flow). 
Then we draw edges where forward data paths are drawn from left to right and backward data paths, representing \textit{backpropagation}, are  drawn as two large backward loops (one for the discriminator and the other for the generator).

\subsubsection*{Visualization of Nodes in Overview Graph}
\label{sec:visual-components}

We visualize the current states of models within the nodes in the graph for users to understand and monitor the training process.

\textbf{Using 2D datasets to promote comprehension.}
One challenge in visualizing this information arises from the difficulty of visualizing a large number of high-dimensional data points.
To tackle this issue,
we decided that we limit our GAN models to generate  two-dimensional data samples,
while GANs often work with high-dimensional image data.
This decision is mainly for helping users easily interpret visualization and focus to understand the internal mechanisms of the models.
As many researchers identified, when designing interactive tools, it is even more desirable to focus on simpler cases~\cite{schweitzer2007interactive}.
Visualization of two-dimensional space is easier for people to understand how data are transformed by the models than that of higher- or one-dimensional spaces:
3D or larger requires dimensionality reduction techniques that add more complexity to users and hinders their understanding.

\vspace{2pt}
Below we describe how we visualize each node.
We show a miniaturized copy of each node's visualization from \autoref{figure:screenshot} for easier referencing.

\begin{wrapfigure}{r}{0.14\columnwidth}
\vspace{-12pt}
\includegraphics[width=\linewidth]{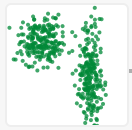}
\vspace{-19pt}
\end{wrapfigure}
\smallskip
\noindent \textbf{Real samples} are what a GAN would like to model.
Each sample, a two-dimensional vector, is represented as a green dot, where its x and y  position represents the values of its two-dimensional data point. 
In this example, two Gaussian distributions exist: on the upper-left, and on the right.

\begin{wrapfigure}{r}{0.14\columnwidth}
\vspace{-12pt}
\includegraphics[width=\linewidth]{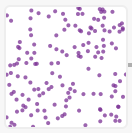}
\vspace{-19pt}
\end{wrapfigure}
\smallskip
\noindent \textbf{Random noise}, an input to the generator,
is a set of random samples.
In \name{}, noise can be either 1D or 2D.
If it is a 1D value, data points are positioned in a line; if a 2D vector (which is default), positioned in a square box, as shown in the small figure on the right.

\begin{wrapfigure}{r}{0.14\columnwidth}
\vspace{-12pt}
\includegraphics[width=\linewidth]{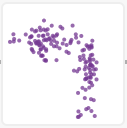}
\vspace{-19pt}
\end{wrapfigure}
\smallskip
\noindent \textbf{Fake samples} are output produced the generator by transforming the random noise.
Like real samples, fake samples are also drawn as dots, but in purple.
For a well-trained GAN, the generated distribution should look indistinguishable from the real samples' distribution.

\begin{wrapfigure}{r}{0.16\columnwidth}
\vspace{-12pt}
\includegraphics[width=\linewidth]{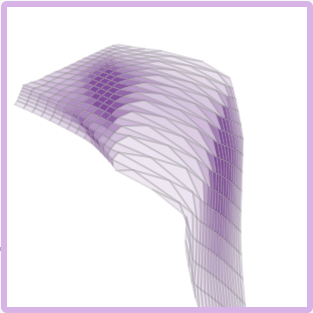}
\vspace{-19pt}
\end{wrapfigure}
\begin{figure}[tb]
 \centering
 \includegraphics[width=\columnwidth]{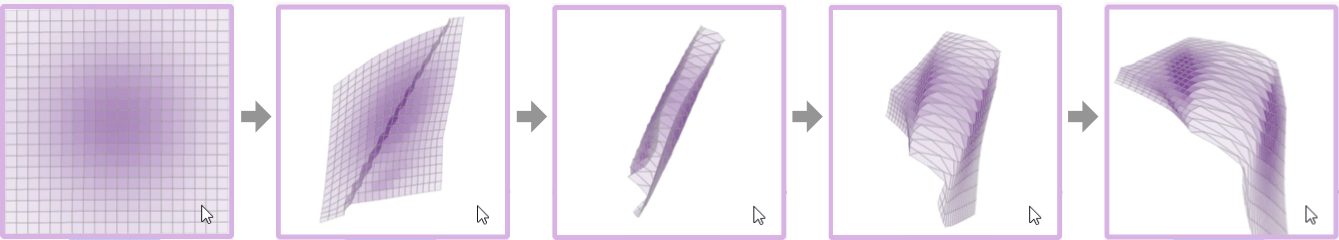}
 \vspace{-5mm}
 \caption{Visualization of generator's transformation.
   When users mouse over the generator node, an animation of the square grid transitioning into a warped version is played.}
 \label{figure:generator-animation}
\end{figure}

\smallskip
\noindent \textbf{Generator}, 
a neural net model,
is a transformation function, $G: \mathbb{R}^2 \rightarrow \mathbb{R}^2$, that maps a 2D data point (i.e., random noise, $\mathbf{z}$) to another 2D data point (i.e., fake sample, $G(\mathbf{z})$).
We visualize the transformed results as a 2D \textbf{manifold}~\cite{olah2014neural}, as in the figure on the right.
To draw this manifold, 
we first create a square grid (e.g., 20x20) for the random noise (see \autoref{figure:generator-animation}, leftmost) where each cell represents a certain noise range 
(e.g., $\{\mathbf{z} = (z_1, z_2) \mid 0.85 \leq z_1 < 0.90 \land 0.10 \leq z_2 < 0.15)\}$).
We color each cell in purple, encode its probability density with opacity (i.e., more opaque means more samples in the cell). 
The generator $G$ transforms the random noise into fake samples by placing them in new locations. 
To determine the transformation for the grid cells, 
we feed each cell's four corners into the generator, which  returns their transformed positions 
forming a quadrangle (e.g., $G(0.85, 0.10) = (0.21, 0.75)$, $G(0.85, 0.15) = (0.24, 0.71)$, ...).
Thus, the whole grid, now consisting of irregular quadrangles, would look like a warped version of the original regular grid.
The density of each (warped) cell has changed.
We calculate its new density by dividing the original density value (in the input noise space) by the \textit{area} of the quadrangle.
Thus, a higher opacity means more samples in smaller space.
Ideally, a very fine-grained manifold will look almost the same as the visualization of the fake samples.
Our visualization technique aligns with the \textit{continuous scatterplots} idea~\cite{bachthaler2008continuous} that generalizes scatterplots to continuous data by computing the density of data samples in the scatterplot  space.
To help users better understand the transformation,
we show an animation of the square grid transitioning into the warped version (see \autoref{figure:generator-animation}), 
when users mouse over the generator node in the overview graph.

\begin{figure}[!b]
 \centering
 \includegraphics[width=0.95\columnwidth]{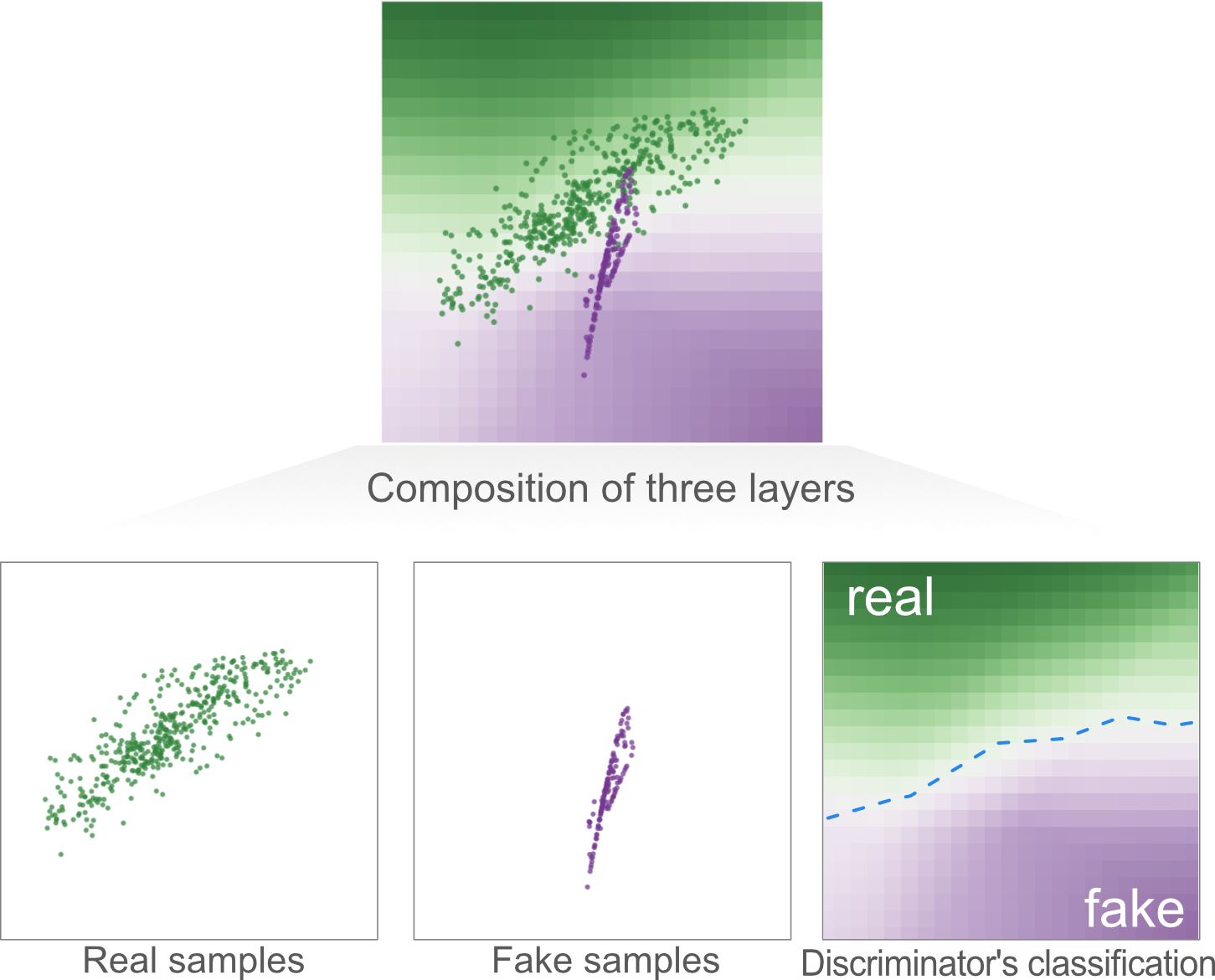}
 \vspace{-2mm}
 \caption{The discriminator's performance can be interpreted through the \textit{layered distributions} view, a composite visualization composed of 3 layers selected by the user:
 \textit{Real samples}, 
 \textit{Fake samples}, 
 and \textit{Discriminator's classification}.
 Here, the discriminator is performing well, since most \textcolor{my_green}{real samples} lies on its classification surface's green region (and 
 \textcolor{my_purple}{fake samples} on purple region).
 }
 \label{figure:layers-discriminator}
\end{figure}

\begin{wrapfigure}{r}{0.16\columnwidth}
\vspace{-12pt}
\includegraphics[width=\linewidth]{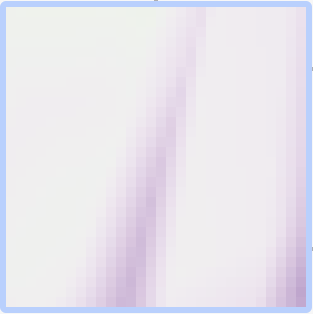}
\vspace{-19pt}
\end{wrapfigure}
\smallskip
\noindent \textbf{Discriminator} is another neural net model, which is a binary classifier, that takes a sample as input and determines whether it is real or fake by producing its prediction score (values from 0 to 1).
We visualize the discriminator using a 2D \textbf{heatmap}, as in TensorFlow Playground~\cite{smilkov2016directmanipulation}.
The background colors of a grid cell encode the prediction values 
(darker green for higher values representing that samples in that region are likely real; darker purple for lower values indicating that samples are likely fake).
As a GAN approaches the optimum, the colors become more gray (as in the above figure), indicating the discriminator cannot distinguish fake examples from the real ones.

\begin{wrapfigure}{r}{0.14\columnwidth}
\vspace{-12pt}
\includegraphics[width=\linewidth]{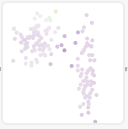}
\vspace{-19pt}
\end{wrapfigure}
\smallskip
\noindent \textbf{Predictions} are outputs from the discriminator. We place real or fake samples at their original positions, but their fill colors now represent prediction scores determined by the discriminator. Darker green indicates it is likely a real sample; darker purple likely a fake sample.
In this example, most samples  are predicted as fake, except for the ones on the upper left.

\begin{wrapfigure}{r}{0.14\columnwidth}
\vspace{-12pt}
\includegraphics[width=\linewidth]{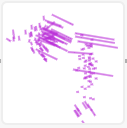}
\vspace{-19pt}
\end{wrapfigure}
\smallskip
\noindent \textbf{Gradients for generator} are computed for each fake sample by backpropagating the generator's loss through the graph.
This snapshot of gradients indicates that how each sample should move to, in order to decrease the loss value. 
As a gradient represents a vector, we visualize it as a line starting from the position of each sample, where length indicates strength.

\begin{figure}[tb]
 \centering
 \includegraphics[width=\columnwidth]{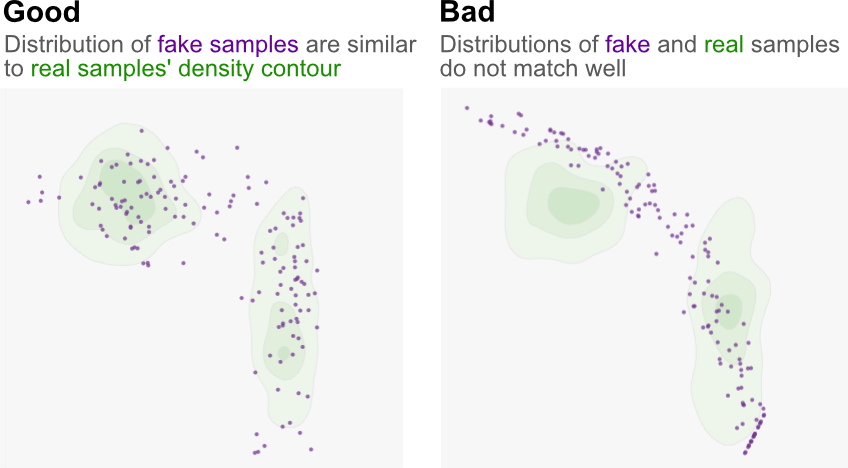}
 \vspace{-5mm}
 \caption{Evaluating how well the distribution of fake samples matches that of real samples by turning on \textcolor{my_green}{real samples' density contour} and \textcolor{my_purple}{fake samples} in the layered distributions view.}
 \label{figure:layers-fake-samples}
\end{figure}

\subsection{Layered Distributions: Visual Analysis of Interplay between Discriminator and Generator}
\label{sec:layered}

In complex models like GANs, it is a key to understanding relationships among several elements of the models.
For example, users may want to check how the distribution of fake samples are similar to those of real samples. 
Although users can perform a side-by-side comparison of the two different nodes on the model overview graph,
this task would be greatly improved when they are overlapped in the same coordinates.

To help visually analyzing relationships among multiple components, we create a \textit{layered distributions} view (\autoref{figure:screenshot} at B) 
that presents a large canvas showing the visual representations of the nodes in the model overview graph as multiple layers.
The layers can be turned on or off using toggle switches.
We do not intend to visualize all layers, 
as it is overwhelming to users and
it is much more effective to include only the useful information for particular tasks.
The view currently supports six layers. 
All layers, except the one for the real samples' density contour, 
are magnified versions of the visual representations of the graph nodes we described in the previous subsection (\autoref{sec:architecture}).
The layers are:

\begin{itemize}[topsep=0mm, itemsep=0mm, parsep=0mm, leftmargin=10mm]

\item Real samples (green dots)

\item Real samples' density contour (see \autoref{figure:layers-fake-samples})

\item Generator transformation manifold 

\item Fake samples (purple dots)

\item Discriminator's classification heatmap 

\item Generator's gradients (pink lines)

\end{itemize}

\vspace{1mm}

\begin{figure}[tb]
 \centering
 \includegraphics[width=0.65\columnwidth]{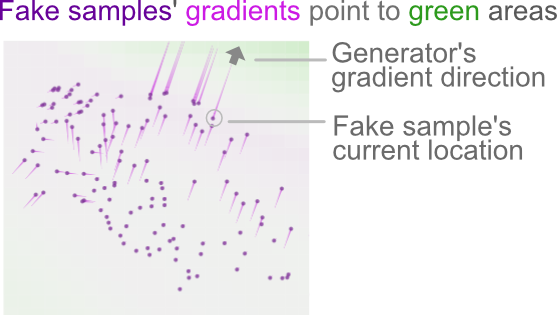}
 \vspace{-2mm}
 \caption{Example of understanding the interplay between discriminator and generator using the layered distributions view.
 \textcolor{my_purple}{Fake samples}' movement directions are indicated by the generator's \textit{gradients} (pink lines), based on those samples' current locations and the discriminator's current classification surface (visualized by background colors).
 }
 \label{figure:layers-guidance}
\end{figure}

\textbf{Useful combinations of layers.}
By selecting which visualizations to be included in the canvas, users can visually analyze the state of the models and the interplay between discriminator and generator, from multiple angles.
We describe three example combinations that support multiple analysis tasks.
First, \autoref{figure:layers-discriminator} illustrates that the discriminator may be visually interpreted
by comparing the samples' positions with grid's background colors.
Here, the discriminator is performing well, as most real and fake samples lie on its classification's green and purple regions, respectively.
The second example in \autoref{figure:layers-fake-samples} illustrates how users may visually evaluate how well the distribution of fake samples matches that of the real samples.
It helps users to determine whether the two distributions are similar or not, which is the main goal of GANs.
The last example in \autoref{figure:layers-guidance} shows how the view can help users understand the interplay between discriminator and generator. 
Fake samples' gradient directions point to the classification's green regions, meaning that
the generator leverages information from the discriminator to make fake samples less distinguishable from the real ones.

\subsection{Metrics: Monitoring Performances}
\label{sec:metrics}

The \textit{metrics view} (\autoref{figure:screenshot} at C) shows a number of line charts that track several metric values changing as the training promises.
\name{} currently provides two classes of metrics.
The first kind is
the loss values of the discriminator and generator,
which are helpful for evaluating submodels and comparing their strengths.
The second kind of metrics is
for evaluating 
how similar the distributions of real and fake samples are.
\name{} provides Kullback-Leibler (KL) and Jensen-Shannon (JS) divergence values~\cite{lin1991divergence,theis2015note}
by discretizing the 2D continuous space (via the grid). 
Formally, the KL divergence value is defined as 
${KL} (P_{\text{real}} || P_{\text{fake}}) = - \sum_i P_{\text{real}}(i) \log \frac{P_{\text{fake}}(i)}{P_{\text{real}}(i)} $,
where $P_{\text{real}}(i)$ is the probability density of the real samples in the $i$-th cell, calculated by dividing the number of  real samples in the $i$-th cell by the total number of real samples; 
$P_{\text{fake}}(i)$ is similarly defined for the fake examples.
We decided to use these measures, among others,
because they are some of the most commonly used approaches for comparing distributions and they do not incur heavy in-browser computation overhead.

\section{Interactive Experimentation}
\label{sec:experimentation}

This section describes how users can interactively experiment with GAN models using \name{}.

\textbf{Basic workflow.}
Clicking the play button, located on the top of the interface, starts running the training of a GAN model and dynamically updates the visualizations of intermediate results every $n$ epochs (a.k.a., iterations).
This helps users keep track of the model's training and examine how they evolve.
Users can pause the training by clicking the pause button (the play button changes to pause button during training).

\subsection{Direct Manipulation of Hyperparameters}
\label{sec:manipulate-options}

\begin{wrapfigure}{r}{0.21\columnwidth}
\vspace{-12pt}
\includegraphics[width=\linewidth]{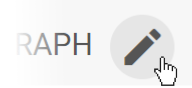}
\vspace{-19pt}
\end{wrapfigure}

\begin{figure}[!tb]
 \centering
 \includegraphics[width=0.8\columnwidth]{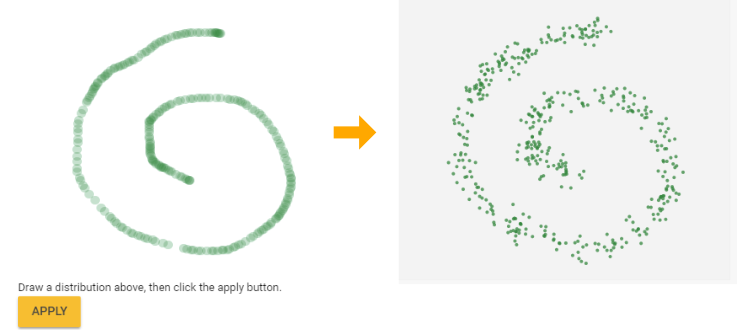}
 \vspace{-3mm}
 \caption{Users can create real samples by drawing their distribution.}
 \label{figure:drawing}
\end{figure}

\name{} is designed for users to directly manipulate model's training as easy as possible.
When users click the editing icon on the right side of the label for the \textit{model overview graph} view,
several up/down buttons or dropdown menus, which controls the model's hyperparameters, are shown (see \autoref{figure:screenshot}).
Each item is located near its relevant submodel or data node for users to easily locate it.
Users can directly change the values using the buttons or dropdown menus, and
the user's actions (e.g., increasing learning rate) are immediately applied to the model training process, except for some of the submodel-specific options (e.g., number of hidden layers), and
the effects of this change will be visualized, as the training further progresses.
This would greatly help users understand how these hyperparameters affect the model training process.
The current available hyperparameters in \name{} include:

\begin{itemize}[topsep=0mm, itemsep=0mm, leftmargin=5mm, parsep=0mm]
\item Number of layers for generator and discriminator
\item Number of neurons in each layer for generator and discriminator
\item Optimizer type (e.g., Stochastic Gradient Descent, Adam) for updating the generator and discriminator 
\item Learning rates for updating the generator and discriminator
\item Loss function (e.g., log loss~\cite{goodfellow2014generative}, least square loss (LS-GAN~\cite{mao2017least}))
\item Number of training runs for discriminator (and generator) for every epoch\footnote{In training of GANs, for every epoch, the discriminator and generator are trained  by turns. Goodfellow et al.~\cite{goodfellow2014generative} suggested that the discriminator can be updated $k$ more times in practice, and \name{} enables to adjust this $k$ value.}
\item Noise dimension (e.g., 1D, 2D) and distribution type (e.g., uniform, Gaussian)
\end{itemize}
\name{} also allows users to pick a distribution of real samples using the drop-down menu that currently implements five examples (e.g., ring).
Users can also specify a new distribution by drawing one on a canvas with brush, as illustrated in \autoref{figure:drawing}.

\subsection{Step-by-Step Model Training at Multiple Levels}
\label{sec:step-all}

\name{} supports \textit{step-by-step} training at multiple levels of abstraction for understanding intricate training dynamics.
The step-by-step execution of systems is one of the useful ways for learners to understand how they work~\cite{shaffer2010algorithm}, however,
training of GANs consists of thousands of iterations and each iteration also consists of several steps (as illustrated in \autoref{figure:step-table}).
To address this problem, 
we decompose the training process into steps in multiple levels: epoch-, submodel-, and component-level.

\subsubsection{Manual Step Execution in Epoch-Level}
\label{sec:step-mode}

\begin{wrapfigure}{r}{0.30\columnwidth}
\vspace{-12pt}
\includegraphics[width=\linewidth]{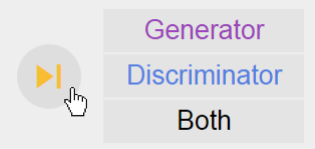}
\vspace{-19pt}
\end{wrapfigure}

Users can train a model for only one epoch, by clicking a button once.
This epoch-level step execution is designed to help users track the training process to see how models update to find the optimum state through iterations.
To use this feature, a user first clicks the step icon on top, which will shows three buttons. The last button (``Both'') represents the training for one epoch. 
We describe the other two buttons' usage next.

\subsubsection{Manual Step Execution in Submodel-Level}

\begin{figure}[!bt]
 \centering
 \includegraphics[width=\columnwidth,trim={6.6cm 12cm 6.0cm 2cm},clip]{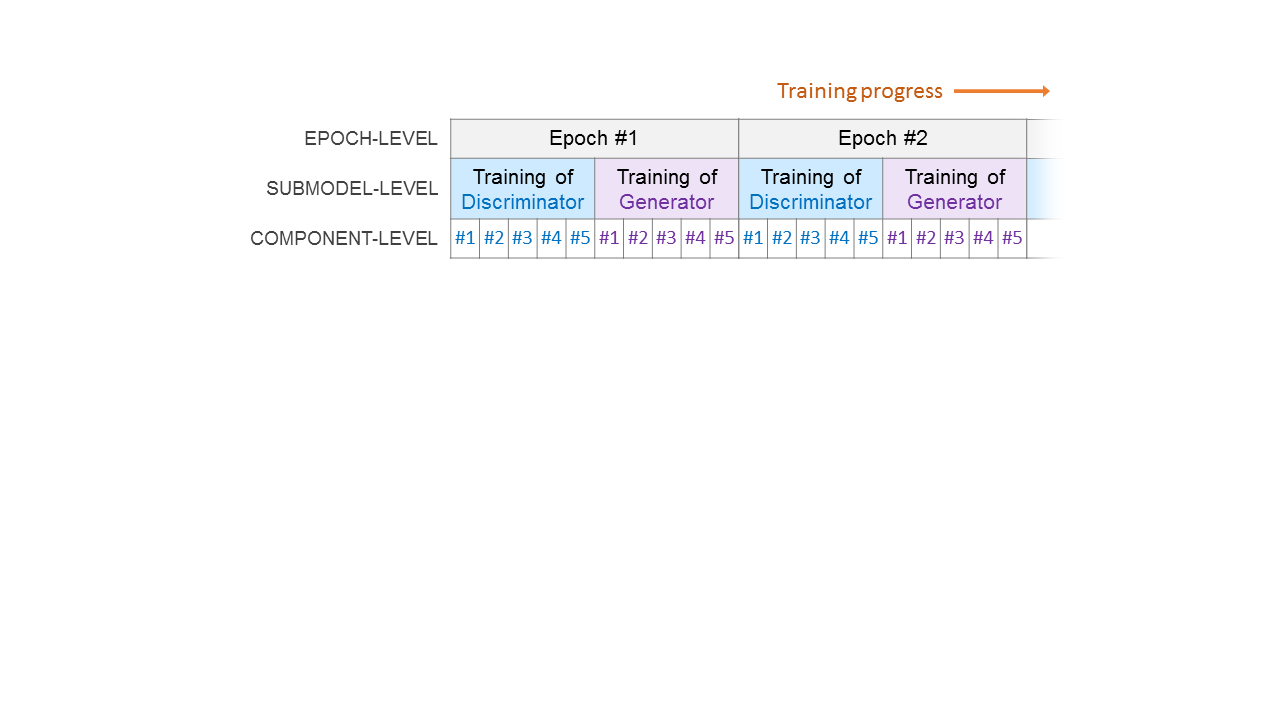}
 \vspace{-5mm}
 \caption{Training typically involves of thousands of epochs (iterations).
 Each epoch includes training both discriminator and generator. \name{} supports step-by-step model training at different abstraction levels.}
 \label{figure:step-table}
\end{figure}

A single epoch consists of 
training of a discriminator and z
generator, as illustrated in \autoref{figure:step-table}.
\name{} allows users to update only the discriminator or generator.
The experimentation of training only one of the two submodels is effective for users to understand how they work differently.
For example, clicking the button for the discriminator changes the background grid while preserving the positions of fake samples. 
On the other hand, clicking the discriminator button moves the fake samples while fixing the background grid.
To use this feature, users click the step icon first, then the three buttons will be shown. The first button is for training the discriminator; the second button is for the generator; and
the last button is for training both submodels.

\subsubsection{Slow-Motion Mode in Component-Level}
\label{sec:slow-mode}

\begin{wrapfigure}{r}{0.10\columnwidth}
\vspace{-12pt}
\includegraphics[width=\linewidth]{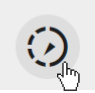}
\vspace{-19pt}
\end{wrapfigure}

\begin{figure*}[!t]
 \centering
 \includegraphics[width=\linewidth]{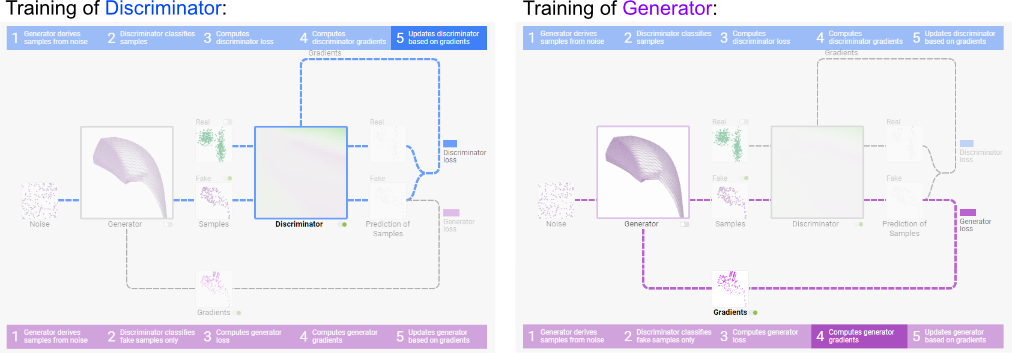}
 \vspace{-5mm}
 \caption{The slow-motion mode slowly executes the model training process in a component level, in a step-by-step fashion. The steps are grouped into two lists, one for discriminator and the other for generator, each consisting of five steps.}
 \label{figure:slow-steps}
\end{figure*}

\name{} also provides the \textit{slow-motion mode},
designed to help novices learn how each component of the model works to make updates within each  epoch.
It works differently from the manual step execution described in the two previous paragraphs.
When users turn on this mode by clicking the icon on top during training, 
it slows down the speed of training.
In addition, two similar lists of five steps are presented: one for updating the discriminator and the other for the generator, as depicted in \autoref{figure:slow-steps}.
The five steps include (1) running the generator; (2) running the discriminator; (3) computing discriminator or generator loss; (4) computing gradients; and (5) updating the discriminator or generator.
For every few seconds, it moves to the next step highlighting the corresponding model components with textual descriptions. 
For example, each of the five steps for the discriminator is highlighted one after another.
At the same time, the whole training loop for the discriminator is also highlighted (i.e., edges colored by blue).
Once the five steps are completed,
it proceeds to the training of the generator, highlighting the training loop for the generator (i.e., purple edges) and executing its five steps.
By following these training paths, users can learn how every component is used in training GANs.

\subsection{Browser-based Implementation for Deployment}
\label{sec:implementation}

\name{} is an open-source, web-based visualization tool.
Anyone can access it using their modern web browsers
without the need for installation or specialized backend.
The demo is currently available at \url{https://poloclub.github.io/ganlab/}.

The tool is implemented in HTML and TypeScript (a typed version of JavaScript) with a few open-source JavaScript libraries: 
\textit{TensorFlow.js}\footnote{TensorFlow.js, \url{https://js.tensorflow.org/}} is 
used for training and running models, which we will elaborate in detail in the next paragraph;
\textit{Polymer}\footnote{Polymer, \url{https://www.polymer-project.org/}} is 
used for building web applications; and
\textit{D3.js}\footnote{D3.js, \url{https://d3js.org/}} 
is used to visualize the model overview graph and layered distributions.
The source code is available in \url{https://github.com/poloclub/ganlab/}.

\textbf{Using \textit{TensorFlow.js} for model building and training.}
\name{} runs locally on user's web browsers
by using \textit{TensorFlow.js Core} (formerly known as \textit{deeplearn.js}),
an in-browser GPU-accelerated deep learning library, developed by Google.
The \textit{TensorFlow.js} library uses WebGL to efficiently perform computation on browsers, required for training deep learning models.
Not only does it enable rapid experimentation of the models,
but also allows us to easily scale up deployment for the public.
While most other implementations of GANs that use Python or other server-side languages would backfire when multiple users train models concurrently,
our GAN models are trained in JavaScript,
which means that that
the models and their visualizations run locally on web browsers,
enabling us to significantly broaden people's access to \name{} for learning about GANs.

\section{Usage Scenarios}
\label{sec:scenario}

This section describes two example usage scenarios for \name{}, demonstrating how it may promote user learning of GANs.
The scenarios highlight:
(1) how beginners may learn key concepts for GANs by experimenting with the tool's visualizations and interactive features (\autoref{ssec:beginners});
(2) how the tool may help practitioners discover advanced inner-workings of GANs, and how it can assist them to interactively attain optimal hyperparameters for reaching equilibrium between submodels (\autoref{ssec:practitioners}).

\subsection{Beginners Learning Concepts and Training Procedure}
\label{ssec:beginners}

Consider Alice, a data scientist at a technology company,
who has basic knowledge about machine learning.
Recently, she has started to learn about deep learning, and a few of the introductory articles she has been reading mention GANs.
Excited about their potential, she wishes to use \name{} to interactively learn GANs.

\begin{figure}[!t]
 \centering
 \includegraphics[width=\columnwidth]{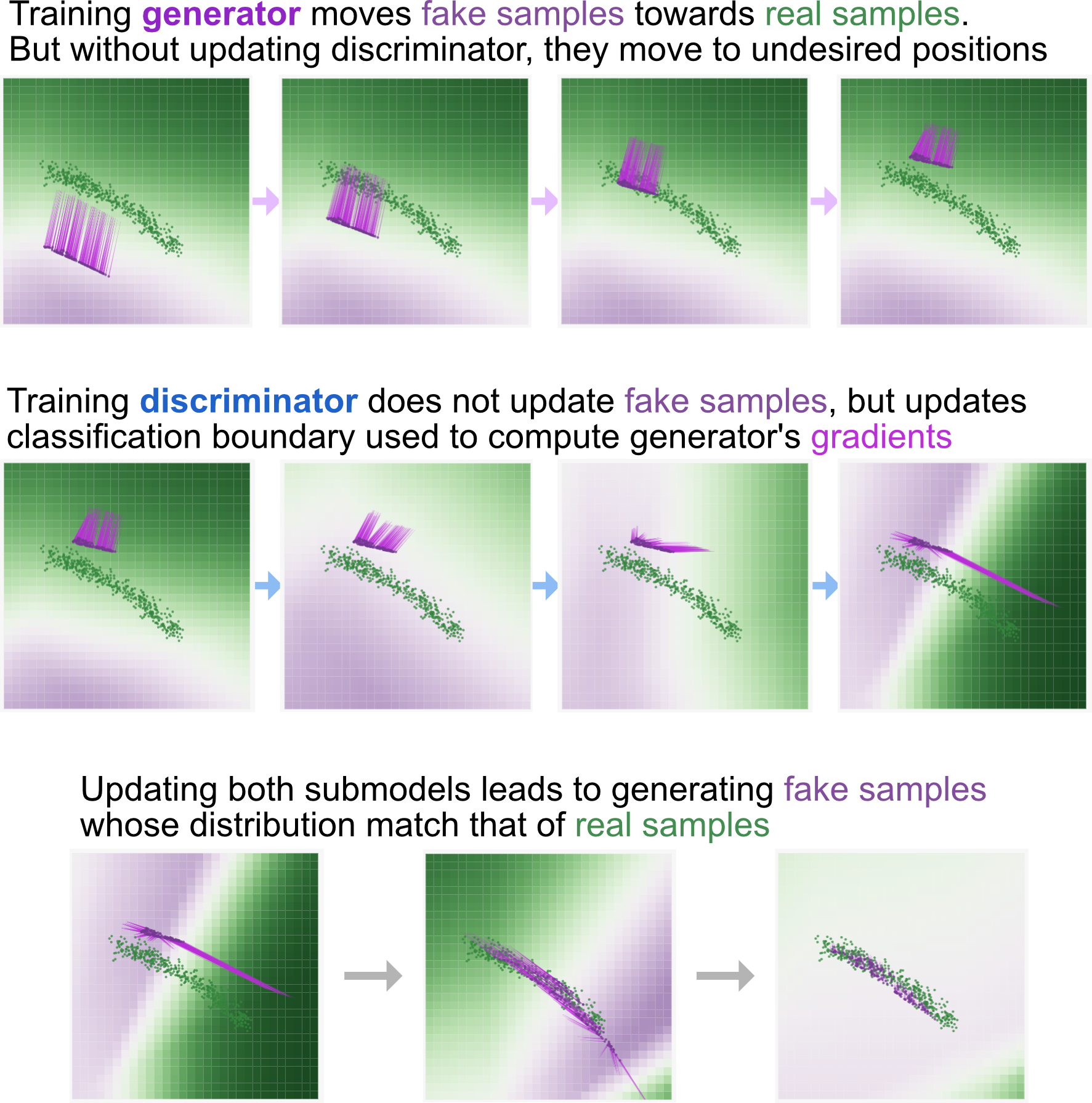}
 \vspace{-2mm}
 \caption{Experimenting with manual step execution, to understand the interplay between discriminator and generator.
  }
 \label{figure:scenario-1}
\end{figure}

\textbf{Becoming familiar with basic workflow.}
When Alice launches \name{} in her web browser,
she sees the \textit{model overview graph}, which looks like a GAN architecture diagram that she has seen in her articles.
By default, \textit{real samples} are drawn from a 2D distribution that resembles a line.
She clicks the \textit{play} button on the tool bar.
During the training, 
the movement of the fake samples in the \textit{layered distribution} view attracts her attention. 
They keep moving towards the real samples.

\textbf{Using the slow-motion mode for tracking the training procedure.}
Alice is aware that
discriminator and generator take turns to train, 
but she is unsure of what that means.
To see how training progresses,
Alice clicks the slow-motion icon (\autoref{sec:slow-mode}) to enter the \textit{slow-motion} training mode, 
which slows down the speed of training, 
and presents two lists of training steps,
one for the discriminator,  and another for the generator (see \autoref{figure:slow-steps}). 
She notices that in for every epoch,
the discriminator is trained first, 
then the generator follows.
The two models' training sequences seem very similar, but she discovers several key differences.
For example, she is able to find that
while discriminator's loss is computed by using both real and fake samples,
only fake samples are used when computing the generator's loss.

\textbf{Understanding the different roles of discriminator and generator with the manual step execution.}
While the slow-motion mode has helped her better understand the steps of the training process,
Alice wonders how the discriminator and generator play a ``game'' to generate data distributions.
To analyze the different effects for the discriminator and the generator,
she would like to experiment with the two submodels using the manual step-by-step execution feature. 
She clicks the button (\autoref{sec:step-mode}) to update the generator.
Her initial clicks cause the fake samples to move towards the real samples,
but as she clicks a few more times, the fake samples ``overshoot,'' no longer matching real samples' distribution (\autoref{figure:scenario-1}, top row).
She now realizes that the fake samples have moved towards regions where the colors of background grid cells are green, 
not directly towards the real samples.
This leads Alice to hypothesize that training the discriminator is necessary for the generator to produce better fake samples. So, she switches to only training the discriminator, which does not reposition the fake samples, 
but the grid colors update (\autoref{figure:scenario-1}, second row) to correct a decision boundary that separates the real and fake samples.
She believes that this new boundary 
helps guide the fake samples towards desirable regions where the real samples are located.
This experiment helps her realize
that updating both submodels is important for generation of better fake samples.
Now she clicks the buttons for updating the discriminator and generator alternatively, 
which successfully creates a fake distribution that matches the real distribution.
That is, the discriminator cannot distinguish between real and fake samples.
(\autoref{figure:scenario-1}, last row).

\subsection{Practitioners Experimenting with Hyperparameters}
\label{ssec:practitioners}

One of \name{}'s key features is the interactive, dynamic training of GANs.
Experimentation using \name{} could provide valuable practical experience in training GAN models even to experts.
Consider Bob, a machine learning engineer at a technology company.

\textbf{Guiding models to find the optimum.}
Bob launches \name{} and starts the training process.
Fake samples quickly move towards real samples.
However,
as the training progresses, he notices that the fake samples oscillate around the real samples. 
Based on his previous experience, he believes this indicates that the \textit{learning rates} may be set too high.
He first decreases the value for the discriminator by using the dropdown menu, but
the amount of oscillation becomes more severe.
By checking the interface, he quickly realizes that there are two learning rates in GANs, so he
reverts its value and decreases the generator's learning rate.
After a few more iterations, the oscillation subsides and the distribution of the fake samples almost matches that for the real samples.
This experimentation helps him understand the importance in balancing the power between the discriminator and generator.

\textbf{Understanding equilibrium between discriminator and generator.}
Bob wonders what would happen if he perturbs the equilibrium between the discriminator and generator. That is, what if either submodel overpowers its complement.
Looking into the model overview graph, he finds that some other hyperparameters also come in matched pairs, such as the number of training loops,
one for the discriminator and the other for the generator.
Originally, both numbers are set to 1 (i.e., the submodels run one training epoch in alternate sequence).
Bob decides to increase discriminator's loop count 3 (i.e., 3 discriminator epochs, followed by 1 generator epoch, followed, and repeat).
To his surprise, this ``unbalanced'' epoch setting (3 vs. 1) causes GAN to converge faster.
Comparing this ``unbalanced'' setting with the original ``balanced'' (1 vs. 1) setting, Bob starts to understand that a more powerful discriminator can indeed accelerate training, 
because a stronger discriminator leads to stronger gradients for the generator, 
which in turns more quickly move the fake samples towards the real distribution, thus faster training convergence.

\begin{figure}[!tb]
 \centering
 \includegraphics[width=0.68\linewidth]{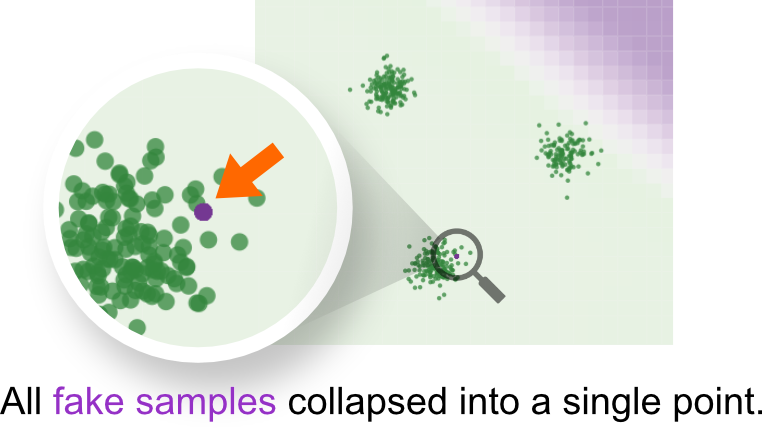}
 \vspace{-2mm}
 \caption{Mode collapse, a common problem in GANs.}
 \label{figure:scenario-2}
\end{figure}

\textbf{Exploring mode collapse.}
Bob would like to train a GAN to work with more complex data distributions.
He picks one distribution that consists of three disjoint dense regions.
He increases the number of layers for both the generator and discriminator, then clicks the play button.
After a few seconds, all fake samples seem to have disappeared, as he can only see real samples. 
He temporarily hides the real samples (by toggling their visibility), thinking that they may be covering the fake samples. 
Then, he realizes that all fake samples have collapsed into a single point (as shown in \autoref{figure:scenario-2}).
He does not know why this happens, and wonders if it is due to his hyperparameter choices.
So he experiments with several other sets of hyperparameters,
and observes the pattern that this happens more often when the generators and discriminators are set to use more layers and neurons.
He consults the literature for possible causes, and learns that this is in fact a well-known problem in GANs, called \textit{mode collapse}, whose exact cause is still an active research topic~\cite{goodfellow2016nips,metz2016unrolled}.
Bob's observation through \name{} motivates him to study new variants of GANs, which may overcome this problem~\cite{goodfellow2016nips,metz2016unrolled}.

\section{Informed Design through Iterations}

The current design of \name{} is the result of 11 months of investigation
and development through many iterations.
Below we share two key lessons learned from our experience.

\begin{figure}[!tb]
 \centering
 \includegraphics[width=0.92\linewidth,trim={0cm 4.1cm 0cm 0cm},clip]{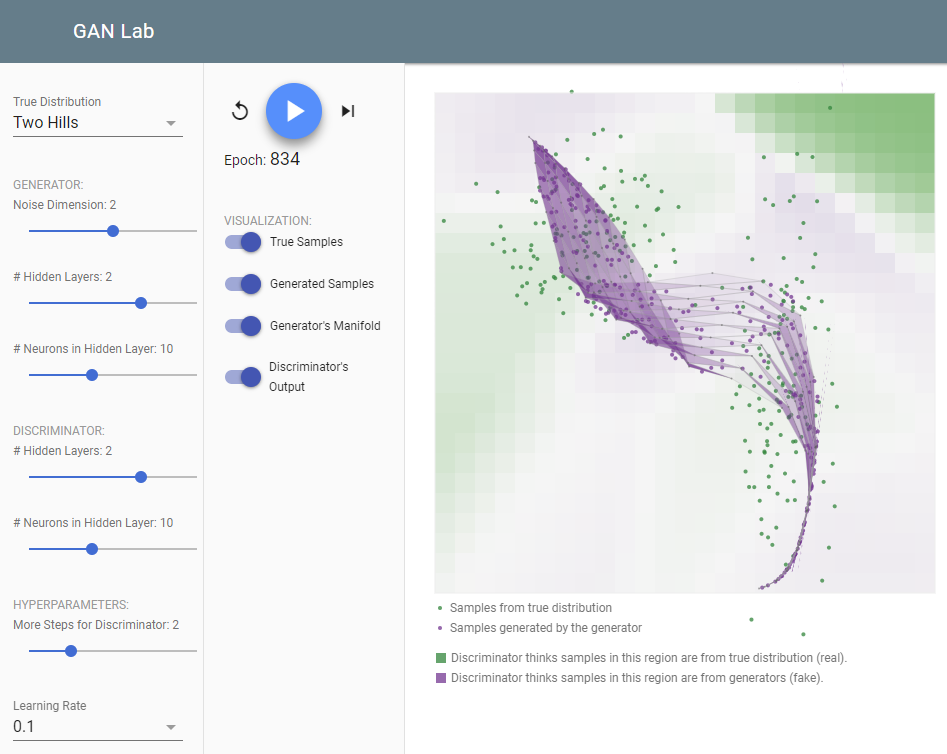}
 \vspace{-2mm}
 \caption{Early design of \name{} did not include  a model overview graph that helps users develop mental models for GANs.}
  \label{figure:early}
\end{figure}

The model overview graph is a crucial and effective feature that helps users develop mental models for GANs.
Our early design (\autoref{figure:early}) did not include the overview graph.
Instead, it displayed a long list of hyperparameters.
While that design had all the necessary features for training GANs interactively,
pilot users,
including machine learning experts, 
commented that the tool was difficult to use and to interpret.
The main reason is that, without an overview, 
users had to develop mental models for GANs (in their heads) to keep track of how the larger number of hyperparameters map to the different model components.
This finding prompted us to add the model overview graph, inspired from common architecture diagrams for GANs,
which helps users build mental models for the training process of GANs~\cite{liu2010mental}.

Animating the generator's transformation (\autoref{figure:generator-animation}) was helpful in helping users interpret the manifold visualization.
Our early version only showed the transformed manifold (e.g., \autoref{figure:generator-animation}, rightmost).
However, many users were puzzled by what they saw because, the manifold could be so severely distorted that they could not tell what its original shape was (a uniform 2D grid), thus they could not make the connection to realize that the manifold visualization was indeed representing the generator's output.
We though about adding text to the interface to explain the manifold, but
as \name{} is intended to be used as a standalone tool, 
we would like to keep the visual design compact, and we wanted to include textual descriptions only when necessary.
Thus, we came up with the idea of visually explaining the transformation as an animated transition, 
which was immediately clear to all users.

\section{Limitations and Future Work}

\indent\indent
\textbf{Transferring user knowledge to higher dimensions.}
Our main decision to use 2D datasets is to promote comprehension~\cite{schweitzer2007interactive}.
Through our tool, with 2D datasets, users can gain important knowledge about the overall training process of GANs, and specific details, such as how model components interact over time, how data flow through components, and how losses are recomputed to iteratively update components.
These important concepts and knowledge are transferable to practical use cases of GANs where higher dimensional data are used (e.g., images).
However, it remains an open research problem whether certain behaviors (e.g., mode collapse) that users may observe when experimenting with 2D datasets would be easily reproducible in higher dimensional datasets,
where the larger number of parameters would lead to more-complex interactions and less-predictable results.
We plan to conduct studies to develop deeper understanding of how and when such correspondence or mismatch may occur.

\textbf{Supporting image data.}
To extend \name{} to support image data, some modifications and optimizations will be needed.
Training on image data is often time consuming.
To speed this up, pre-trained models may be provided to  users so they can skip the earlier training steps.
As for visual design,
projection methods (e.g., t-SNE) may be used to replace some views in \name{} to visualize the distribution of generated image samples~\cite{wang2018ganviz}.

\textbf{Speed and scalability.}
\name{} leverages TensorFlow.js to accelerate GAN training for browser-based deployment.
For models with many parameters, this can be time consuming.
In the short term, we believe rapid advances in JavaScript and hardware will shorten this by a good amount.
A longer-term challenge to overcome is browsers' inability to render visualization and perform computation at the same time (i.e., single-threaded).
Developers need to strike a good balance in planning and interleaving these actions, to maximize model computation speed and visual responsiveness.

\textbf{Supporting more GAN variants.}
While \name{} currently implements a few different loss functions, other GAN variants exist~\cite{2017ganzoo}.
Through open-sourcing \name{}, we look forward to seeing the community to build on \name{} to implement more variants,
enabling users to interactively and visually compare them, easing the challenges in evaluating GANs~\cite{goodfellow2016nips}.
Some variants may require minor design changes of the interface (e.g., adding new nodes to overview graph).

\textbf{In-depth evaluation of educational benefits.}
Longitudinal studies of \name{} will help us better  understand how it helps with learning of GANs.
It would be particularly valuable to investigate how different types of users (e.g., students, practitioners, and researchers) would benefit from the tool.

\acknowledgments{
We thank Shan Carter, Daniel Smilkov, Google Big Picture Group and People + AI Research (PAIR), Georgia Tech Visualization Lab, and the anonymous reviewers for their feedback. 
This work was supported in part by NSF grants IIS-1563816, CNS-1704701, and TWC-1526254.
}

\bibliographystyle{abbrv-doi-hyperref-narrow}

\bibliography{references}
\end{document}